\documentclass[conference]{IEEEtran}
\IEEEoverridecommandlockouts
\usepackage{algorithm}
\usepackage{algpseudocode}
\usepackage{amsmath,amssymb,amsfonts}
\usepackage{bbm}
\usepackage{cite}
\usepackage{graphicx}
\usepackage{nicefrac}
\usepackage{url}

\algnewcommand\algorithmicinput{\textbf{Input:}}
\algnewcommand\INPUT{\item[\algorithmicinput]}
\algnewcommand\algorithmicoutput{\textbf{Output:}}
\algnewcommand\OUTPUT{\item[\algorithmicoutput]}
\DeclareMathOperator*{\argmin}{arg\,min}

\def\BibTeX{{\rm B\kern-.05em{\sc i\kern-.025em b}\kern-.08em
    T\kern-.1667em\lower.7ex\hbox{E}\kern-.125emX}}
\begin{document}

\newcommand{\supp}[1]{\mathrm{supp}({#1})}
\newcommand{\Fig}[1]{Fig.~\ref{#1}}

\title{On the Error-Reducing Properties\\of Superposition Codes
\thanks{The research was carried out at Skolkovo Institute of Science and Technology and supported by the Russian Science Foundation (project no. 23-11-00340), \protect\url{https://rscf.ru/en/project/23-11-00340/}}
}

\author{\IEEEauthorblockN{Kirill Andreev, Pavel Rybin, Alexey Frolov}
\IEEEauthorblockA{\textit{Center for Next Generation Wireless and IoT (NGW)} \\
\textit{Skolkovo Institute of Science and Technology}, Moscow, Russia \\
k.andreev@skoltech.ru, p.rybin@skoltech.ru, al.frolov@skoltech.ru}
}

\maketitle

\begin{abstract}
Next-generation wireless communication systems impose much stricter requirements for transmission rate, latency, and reliability. The peak data rate of 6G networks should be no less than 1 Tb/s, which is comparable to existing long-haul optical transport networks. It is believed that using long error-correcting codes (ECC) with soft-decision decoding (SDD) is not feasible in this case due to the resulting high power consumption. On the other hand, ECC with hard-decision decoding (HDD) suffers from significant performance degradation. In this paper, we consider a concatenated solution consisting of an outer long HDD code and an inner short SDD code. The latter code is a crucial component of the system and the focus of our research. Due to its short length, the code cannot correct all errors, but it is designed to minimize the number of errors. Such codes are known as error-reducing codes. We investigate the error-reducing properties of superposition codes. Initially, we explore sparse regression codes (SPARCs) with Gaussian signals. This approach outperforms error-reducing binary LDPC codes optimized by Barakatain, et al. (2018) in terms of performance but faces limitations in practical applicability due to high implementation complexity. Subsequently, we propose an LDPC-based superposition code scheme with low-complexity soft successive interference cancellation (SIC) decoding. This scheme demonstrates comparable performance to SPARCs while maintaining manageable complexity. Numerical results were obtained for inner codes with an overhead (OH) of 8.24\% within a concatenated scheme (15\% OH) with an outer hard-decision decoded staircase code (6.25\% OH).
\end{abstract}


\section{Introduction}

With the ongoing development and widespread adoption of cutting-edge communication systems such as 5G, the demands for speed and reliability are continuously increasing. Consequently, it becomes crucial to explore long high-rate codes as a means to address these challenges. However, modern code constructions with soft-decision decoding (SDD) result in high power consumption and significant decoding delay, which complicates their implementation on modern hardware platforms. On the other hand, code constructions with hard decision decoding (HDD) have limited error-correcting capabilities and fail to meet the requirements. A viable solution lies in combining soft and hard approaches.

We consider a concatenated code construction with a powerful outer code and a high-rate inner code to minimize the number of errors. Our focus is on scenarios where the inner code cannot correct all errors due to its high rate and short length. These codes, known as \textit{error-reducing} codes, act as adapters to enhance the channel quality. The concept of error-reducing codes was first introduced in~\cite{Spielman1996erc} to construct asymptotically good codes with linear complexity encoding and decoding algorithms. Furthermore, the error-reducing problem can be viewed as a specific instance of joint source-channel coding (JSCC)~\cite{Polyanskiy2019jscc, Polyanskiy2018jscc, Zamir2006jscc, Verdu2013jscc}. In practice, short Hamming codes with simple and fast Chase decoding are commonly used as inner error-reducing codes \cite{Udo2011hamming, Zhongfeng2018hamming}. However, such schemes require the presence of long interleavers due to the "bad" output distribution and have limited error-reducing capabilities. Motivated by this scenario, in \cite{Polyanskiy2020graceful, Polyanskiy2019ldmg}, the authors showed that non-linear codes can achieve a lower bit error rate (BER) than any linear code in the error-reducing regime for the binary erasure channel (BEC). They also proposed a construction called low-density majority codes (LDMC), which proved to be superior to any linear code as a code that reduces the number of errors. However, it is worth noting that the performance of LDMC codes is not as good in the additive white Gaussian noise (AWGN) channel \cite{Balitskiy2021}. Another approach to constructing error-reducing codes was proposed in \cite{Kschischang2017LDGMStairCase}, where the authors considered a concatenation of staircase codes and linear low-density generator matrix (LDGM) codes. This scheme was later improved in \cite{Kschischang2018LDPCStairCase}, where LDGM codes were replaced with carefully constructed LDPC codes. The key idea of the latter scheme is to achieve the required bit error rate (BER) through unequal protection of bits. However, it is important to highlight the primary issue with the schemes presented in \cite{Kschischang2017LDGMStairCase, Kschischang2018LDPCStairCase}: the length of LDGM and LDPC codes was chosen to be $20000$ bits, which is excessively large for the desired throughput.

Let us delve into the solution from~\cite{Kschischang2018LDPCStairCase} in more detail. We define $\supp{\mathbf{v}}$ as the support of the vector $\mathbf{v}$ or the set consisting of all indices corresponding to nonzero entries in $\mathbf{v}$. Furthermore, let $\mathcal{V} = \bigcup\nolimits_{i} \mathbf{v}_i$, then $\supp{\mathcal{V}} = \bigcup\nolimits_{i} \supp{\mathbf{v}_i}$. The transmission scheme from \cite{Kschischang2018LDPCStairCase} can be presented as a superposition coding $\mathcal{C} = \mathcal{C}_1 + \mathcal{C}_2$, where $\supp{\mathcal{C}_1} \bigcap \supp{\mathcal{C}_2} = \emptyset$. We denote $I_1$ as the support of $\mathcal{C}_1$, $I_2$ as the support of $\mathcal{C}2$, and $\mathcal{C}_\mathcal{I}$ as the restriction of the code $\mathcal{C}$ to the coordinates in the set $\mathcal{I}$. The codes $\mathcal{C}_1$ and $\mathcal{C}_2$ are selected such that $\left( \mathcal{C}_1 \right)_{\mathcal{I}_1}$ represents a BPSK-modulated LDPC code designed for error correction, while $\left( \mathcal{C}_2 \right)_{\mathcal{I}_2}$ is a BPSK-modulated trivial (rate $1$) code.

In this paper, we extend the approach of \cite{Kschischang2018LDPCStairCase} and consider the superposition of $L$ codes with intersecting supports, which naturally leads to the so-called sparse regression codes (SPARCs). Sparse regression codes are currently being actively investigated for single and multi-user communication scenarios (see~\cite{Li2020SparcsMAC, Chamberland2023sparcs} for more details). We find that these codes significantly improve error-reducing capabilities at the expense of complexity. Therefore, we propose low-complexity schemes that utilize non-orthogonal superposition of two BPSK-modulated LDPC codes with different power levels. Several decoding algorithms, such as successive interference cancellation (SIC) and soft SIC decoders, are considered. We find that this scheme provides a trade-off between performance and complexity. Additionally, it is worth noting that all of the proposed schemes utilize higher modulation orders (compared to BPSK), for example, the latter $2$-LDPC scheme uses a variant of PAM-4 modulation.

To compare the solutions described above, we utilize the following scenario. An outer code with HDD is employed, namely the Staircase code with 6.25\% overhead (OH, that is defined as $(1/R - 1)\times 100\%$, where $R$ is the coding rate), as described in \cite{Kschischang2014Staircase}. This code is known to achieve a target output bit error level (BER) of $10^{-15}$ (after decoding) at an input BER of $4.7 \times 10^{-3}$, which corresponds to the desired output BER for the inner code. The main focus of this work is on the energy efficiency of the inner code with an overhead of 8.24\%, as it directly influences the energy efficiency of the concatenated scheme, which has an overall OH of 15\%.

\section{Preliminaries}

Our goal is to transmit an information vector $\mathbf{u} = (u_1, \ldots, u_k) \in \left\{0,1\right\}^k$ with given output BER (distortion level) through the noisy channel. To achieve this, we utilize the encoder function $f: \left\{0,1\right\} \to \mathbb{R}^n$ to obtain the codeword $\mathbf{c} = \left(c_1, \ldots, c_n\right) \in \mathbb{R}^n$, which is then transmitted through the channel while satisfying the power constraint $P$. The channel introduces noise, resulting in the observation:
\begin{equation}
\label{eq:channel}
\mathbf{y} = \mathbf{c} + \mathbf{w}, \quad \mathbf{w} \sim \mathcal{N}\left(0, \mathbf{I}_n\right).
\end{equation}
Upon observing the distorted information, the decoder $g$ estimates the bits sent, i.e., calculates $\hat{\mathbf{u}} = g(\mathbf{y})$. We measure the quality of the decoder using the data bit error rate (BER):
$$
P_b = \frac{1}{k} \sum\limits_{i = 1}^k \mathbb{P}\left[u_i \neq \hat{u}_i \right].
$$
We require $P_b \leq D$, where $D$ represents the desired BER or distortion level. Thus, we can formulate the codes error-reducing property, that is under consideration in this paper:
$$
\mathbb{P}\left[ g(\mathbf{y}) \in \mathbb{B}_{k}\left(\mathbf{u}, Dk \right)  \right] \geq 1 - \epsilon,\;\forall \mathbf{u} \in \left\{0,1 \right\}^{k}
$$
where $\mathbb{B}_{k}\left(\mathbf{u}, Dk \right)$ is a $k$-dimensional ball with center in $\mathbf{u}$ and radius $Dk$, and $\epsilon > 0$ is a small positive number.

In other words, it is required that nearest codewords to the given codeword $f\left( \mathbf{u} \right)$ correspond to the information words inside the ball $\mathbb{B}_{k}\left(\mathbf{u}, Dk \right)$ with high probability. 

So, Our goal is to minimize SNR such that $P_{b} \leq D$ for given $D$. Given the SNR, we consider $E_b / N_0 = nP / 2k$ and $E_s/N_0 = P/2$ (to align with previous works).

We define the frame error rate (FER) as $P_f=\mathbb{P}\left[\mathbf{u} \neq \hat{\mathbf{u}}\right]$.

\section{SPARCs approach}

%Sparse regression codes (SPARCs) are known to achieve the capacity of the additive white Gaussian noise (AWGN) channel~\cite{BarronSparcs2019}.
SPARCs are characterized by a \emph{design matrix} $\mathbf{A}$ that has dimensions $n\times ML$, where $n$ is the number of channel uses. In this paper, we consider the matrix $\mathbf{A}$ with elements $\mathbf{A}{i,j}\overset{\text{i.i.d}}{\sim}\mathcal{N}\left(0, \nicefrac{1}{n}\right)$. The columns of $\mathbf{A}$ are divided into $L$ sections, each containing $M$ columns. The transmitted codeword is a linear combination of $L$ columns, with one column selected from each section. This set can be represented as a set of nonzero elements of a sparse vector $\mathbf{b}\in \mathbb{R}^{ML}$. Let $\text{sec}\left(\ell\right)$ be the set of indices within section $\ell$. The $\ell$-th set of $\log_2M$ bits from the information message define the position of the nonzero entry of $\mathbf{b}_j$, where $j\in\text{sec}\left(\ell\right)$ within each section $\ell$. Using this approach, a code of length $n$ (channel uses) with number of information bits $k=\log_{2}M \cdot L$ and rate $R = \nicefrac{k}{n}$ can be constructed. Finally, let us introduce the per-section power allocation scheme $\left\{P_\ell\right\}$ that satisfies the total power constraint $P = \sum_{\ell=1}^{L}P_\ell$. To assign the power $P_\ell$ to the $\ell$-th section, one needs to assign the amplitudes $\sqrt{nP_\ell}$ to the nonzero entries of $\mathbf{b}$ corresponding to section $\ell$. The resulting codeword is $\mathbf{c} = \mathbf{Ab}$. The received signal is given by~\eqref{eq:channel}.
% \begin{equation}
% \label{eq:sparcs_rx}
% \mathbf{y} = \mathbf{A} \mathbf{b} + \mathbf{w}, \quad \mathbf{w} \sim \mathcal{N}\left(0, \mathbf{I}_n\right).
% \end{equation}

Decoding problem is equivalent to finding an estimate $\hat{\mathbf{b}}$ of the transmitted vector $\mathbf{b}$ given the design matrix $\mathbf{A}$ and a received sequence $\mathbf{y}$. This can be formulated as a sparse linear regression problem. LASSO provides a sparse solution to the linear regression problem, where the parameter $\lambda$ controls the trade-off between sparsity and residual error.
\begin{equation}
\label{eq:lasso}
\hat{\mathbf{b}} = 
\argmin_{\hat{\mathbf{b}} \in \mathbb{R}^N}\left\|\mathbf{y} - \mathbf{A}\hat{\mathbf{b}}\right\|^2_2 + \lambda \left\|\hat{\mathbf{b}}\right\|_1
\end{equation}

The problem mentioned above is also known as the compressed sensing (CS) problem~\cite{Donoho2006CS}. Approximate message passing~\cite{Donoho2010AMP} is a low-complexity approximation of loopy belief propagation on dense graphs. This approximation takes advantage of the Gaussian property of the design matrix. Furthermore, AMP converges to the LASSO solution in the presence of noise~\cite{Montanari2012Lasso}.

\begin{algorithm}
\caption{Approximate message passing for SPARCs\label{alg:AMP}}
\begin{algorithmic}
\INPUT{Received sequence $\mathbf{y}$, design matrix $\mathbf{A}$, power allocation $P_\ell$, $\ell \in 1, \ldots, L$}
\State Initial step: $\mathbf{b}^0 = 0$ and $\mathbf{z}^{-1} = 0$
 \For {$t \in 1, \ldots, T$} \Comment Iterative decoding 
\State Calculate the residual
$$
\mathbf{z}^t = \mathbf{y} - \mathbf{A}\mathbf{b}^t + \underbrace{\frac{\mathbf{z}^{t-1}}{\tau^2_{t - 1}}\left(P - \frac{\left\|\mathbf{b}^t\right\|^2}{n}\right)}_{\text{Onsager term}}
$$
\State Calculate the test statistic:
$
\mathbf{s}^t = \mathbf{A}^T\mathbf{z}^t + \mathbf{b}^t
$
\State Apply non-linear term~\eqref{eq:amp_nonlin}:
$
\mathbf{b}^{t + 1} = \eta_t\left(\mathbf{s}^t\right)
$
\State Estimate the residual variance:
$
\tau_t^2 = {\left\|\mathbf{z}^t\right\|^2}/{n}
$
\EndFor
\end{algorithmic}
\end{algorithm}

Let us highlight the main components of AMP (see Algorithm~\ref{alg:AMP}). The first component is the residual vector $\mathbf{z}$, which represents the difference between the received sequence and the estimated noiseless signal. The second component is the Onsager term, which is designed to asymptotically eliminate the correlation between $\mathbf{A}$ and $\left(\hat{\mathbf{b}}^t - \mathbf{b}\right)$. Finally, there is a component-wise nonlinear function $\eta\left(\cdot\right)$, which in the original AMP algorithm is a thresholding function. For SPARCs decoding, this term needs to be modified as follows. First, let us define the set of valid message vectors $\mathbf{b}$ as $\mathcal{B}_{M, L}$. The SPARCs decoding problem differs from the original LASSO problem~\eqref{eq:lasso} as follows.
\begin{equation}
\label{eq:sparcs_ml}
\hat{\mathbf{b}}_{\text{opt}} = \argmin_{\hat{\mathbf{b}} \in \mathcal{B}_{M, L}}\left\|\mathbf{y} - \mathbf{A}\hat{\mathbf{b}}\right\|^2
\end{equation}
This additional constraint $\hat{\mathbf{b}} \in \mathcal{B}_{M, L}$ results in the following form of nonlinear AMP term~\cite{BarronSparcs2019}:
\begin{equation}
\label{eq:amp_nonlin}
\eta_i^t\left(\mathbf{s}^t\right) = \sqrt{nP_\ell}\frac{e^{\mathbf{s}_i^t\sqrt{nP_\ell} / \tau_t^2}}{\sum_{j \in\text{sec}\left(\ell\right)}e^{\mathbf{s}_j^t\sqrt{nP_\ell}/\tau_t^2}}, \quad i \in \text{sec}\left(\ell\right)
\end{equation}
The final note on the SPARCs decoding algorithm is a residual variance estimation. One can use the state evolution equations~\cite{BarronSparcs2019} to estimate $\tau_t^2$, or (as mentioned in Algorithm~\ref{alg:AMP}) calculate it directly from the residual vector $\mathbf{z}^t$.

Sparse regression codes have been shown to exhibit strong error-correcting properties~\cite{BarronSparcs2019} when employed with varying transmit powers allocated to different sections (to enhance successive cancellation) and in conjunction with an outer code capable of correcting a small number of errors after the AMP decoding. Our objective is to utilize SPARCs in the error-reducing regime. Through simulations, we have observed that a non-uniform power allocation scheme is inefficient, and thus we consider a uniform power allocation scheme with $P_\ell = P / L$. Furthermore, it is worth noting that the target BER is typically achieved at signal-to-noise ratios (SNRs) where the FER is equal to one. %Consequently, employing a high-rate outer code becomes meaningless in this scenario.

\section{LDPC-based superposition codes approach}

As demonstrated in Section~\ref{sec:results}, the error-reducing capabilities of SPARCs are highly influenced by the section size. To achieve improved performance, it is necessary to increase the section size $M$, which, in turn, leads to a significant increase in the size of the design matrix $\mathbf{A}$. While using Hadamard design matrices~\cite{Chamberland2023sparcs} eliminates the need to store $\mathbf{A}$ in memory, the matrix-vector multiplication operation will still incur a substantial time cost. In this section, we attempt to mitigate this issue by testing a reduced number of sections through the replacement of $L$ random codes with practically decodable ones. We expect to observe performance improvements resulting from the reduction in the number of sections. However, it is important to acknowledge that this performance gain might be compromised due to the transition from random codes to practically decodable LDPC codes.

Consider the transmitted codeword $\mathbf{c}$ given by
$$
\mathbf{c} = \sqrt{P_1}f_1(m_1) + \sqrt{P_2}f_2(m_2), \quad f_\ell: \left\{0, 1\right\}^{k_\ell}\mapsto\left\{\pm 1\right\}^n,
$$
where $m = m_1 \oplus m_2$ represents the information message split into $m_1$ and $m_2$ with $k_1$ and $k_2$ bits, respectively, such that $k_1 + k_2 = k$. Here, $f_\ell(\cdot)$ denotes the encoding and BPSK modulation function for $\ell \in \{1, 2\}$. The choice of LDPC codes, the allocation of information bits between $k_1$ and $k_2$, and the power allocation scheme $\left\{P_\ell\right\}$ are parameters to be optimized, and we explore this optimization in Section~\ref{sec:results}. We focus on $L=2$ codes in this work to simplify the parameter optimization process, as increasing the number of code components would result in a significantly higher number of parameters to optimize.
The decoding algorithm is inspired by the treat interference as noise (TIN) decoder followed by soft SIC proposed in \cite{Wang2019, Ping2004}. The steps can be iteratively repeated, leading to an iterative soft SIC algorithm described in Algorithm~\ref{alg:SIC}. The interference vector $\boldsymbol{\rho}_\ell$ is estimated using the output log-likelihood ratios (LLRs) of the $j$-th LDPC decoder, where $j \in \{1, \ldots, L\} \setminus \ell$, and the interference power $\mathcal{P}_\ell$ is used to re-calculate the signal-to-interference-plus-noise ratio (SINR) with the expectation $\mathbb{E}[1 - \tanh^2(\boldsymbol{\gamma}_j / 2)]$ taken over all channel uses.

\begin{algorithm}
\caption{Soft interference cancellation decoder\label{alg:SIC}}
\begin{algorithmic}
\State $\boldsymbol{\gamma}_\ell \gets 0$, for $\ell \in 1, \ldots, L$ \Comment Initialize  LLRs for each code
\For {$\text{outer iterations}$}
\For {$\ell \in 1, \ldots, L$} \Comment Try decoding each LDPC code
\State\Comment Estimate interference signal $\boldsymbol{\rho}_\ell$ and its power $\mathcal{P}_\ell$:
\State $\boldsymbol{\rho}_\ell = \sum_{j = \left\{1, \ldots, L\right\}\setminus \ell}\sqrt{P_j}\tanh{\left(\boldsymbol{\gamma}_j / 2\right)}$
\State $\mathcal{P}_\ell = \sum_{j = \left\{1, \ldots, L\right\}\setminus \ell}P_j\mathbb{E}\left[1 - \tanh^2{\left(\boldsymbol{\gamma}_j / 2\right)}\right]$
\State $\text{SINR} = P_\ell / \left(N_0 + \mathcal{P}_\ell \right)$ \Comment Noise power $N_0$
\State $\boldsymbol{\gamma}_\ell^{(in)} = 2 \left(\mathbf{y} - \boldsymbol{\rho}_\ell\right)\times\text{SINR} / {\sqrt{P_\ell}}$ \Comment Input LLRs
\State $\hat{\boldsymbol{\gamma}}_\ell \gets \text{decode}_\ell\left(\boldsymbol{\gamma}_\ell^{(in)}\right)$ \Comment $\ell$-th LDPC decoder
\EndFor
\State ${\boldsymbol{\gamma}}_\ell \gets \hat{\boldsymbol{\gamma}}_\ell $ for $\ell \in 1, \ldots, L$ \Comment Update output LLRs
\EndFor
\end{algorithmic}
\end{algorithm}

At the initial step, the decoder has no knowledge about the interference. Assuming the power allocations $P_\ell$ are known, one can perform BPSK demodulation on the received sequence $\mathbf{y}$ while treating one of the two codewords as interference. The demodulation process results in LLR vectors $\boldsymbol{\gamma}_\ell^{(in)}$. These LLR vectors are then passed to the LDPC decoding algorithms. Given the output LLRs $\hat{\boldsymbol{\gamma}}$, one can derive the expected interference $\boldsymbol{\rho}$ as $\tanh(\hat{\boldsymbol{\gamma}}_\ell / 2)$ \cite{Ping2004}. Additionally, the interference power $\mathcal{P}_\ell$ can be estimated, and it is used to recompute the input LLRs for each decoder.

\section{Numerical results}
\label{sec:results}

\begin{table}
\caption{Reference scenario\label{tab:sim_params}}
\begin{tabular}{l|c}
The number of information bits & $k = 5540$ \\ 
The number of real channel uses & $n = 6000$ \\ 
Target bit error rate & $D = 4.7\times 10^{-3}$ \\ 
Reference signal-to-noise ratio, $[\text{dB}]$ & $3.39$ \\
Modulation & BPSK 
\end{tabular}
\end{table}

\begin{figure}
\includegraphics{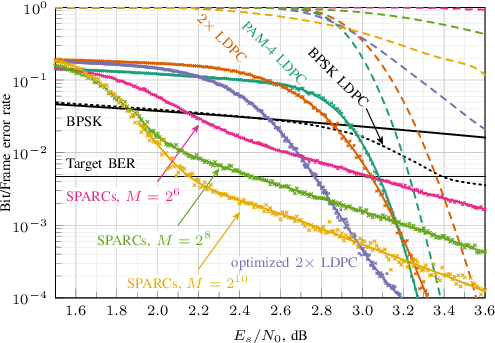}
\caption{BER and FER for different solutions. Frame error rate values are presented with dashed lines. For the bit error rates, the scatter points correspond to actual values, and the lines correspond to a polynomial regression fit with the Bernoulli loss function. The dotted line corresponds to a single LDPC setup optimized for error-reducing.\label{fig:sim_results}}
\end{figure}

%Our reference point for BPSK-modulated and LDPC-encoded sequences is $3.49$ dB (see Table~\ref{tab:sim_params}). Given the SNR, we consider $E_b / N_0 = nP / 2k$, where $P$ is the transmit power, and $E_s/N_0 = P/2$ (to achieve the same uncoded BER in both real- and complex-valued channels).

The numerical results are presented in~\Fig{fig:sim_results}. The solid black line corresponds to the uncoded BPSK BER. The thin black line represents the target BER $D$. The dotted line corresponds to the BER-optimal LDPC code with parameters taken from Table~\ref{tab:sim_params} and labeled as ``BPSK LDPC'' (reference point $3.39$ dB). Note that the BER-optimal LDPC code does not increase the uncoded BPSK BER. This is achieved by replacing the decoded bits with the channel output in the case of decoding failure. For the proposed schemes, we show FER curves in addition to BER curves and do not use the mentioned trick. The colored lines represent different setups, with solid lines denoting BER and dashed lines denoting FER. To find the exact points of intersection with the target BER line, we use polynomial regression with a Bernoulli loss function.

We consider a sparse regression code with $k$ and $n$ taken from Table~\ref{tab:sim_params}. We have simulated SPARCs for section sizes $M=2^6, 2^8, 2^{10}$ (shown by magenta, green, and yellow lines in~\Fig{fig:sim_results}). We use Algorithm~\ref{alg:AMP} with a fixed number of iterations $T=50$. All solutions exhibit an error floor, but the target BER $D$ can be achieved at a much lower SNR compared to the original BPSK-modulated LDPC code with a rate of $R=0.923$. The best-performing solution has a section size of $M=2^{10}$, achieving the target BER at $2.16$ dB (see~\Fig{fig:sim_results}, green curve). $M=2^{10}$ was the maximum section size that we could simulate.

For LDPC superpositioning, we use $20$ iterations of sum-product decoding at each of the $20$ outer iterations. All LDPC codes are randomly interleaved to randomize the interference. We started our optimization by finding the optimal rate and power splits, and then performed an optimization of the LDPC codes at the second step.

\begin{figure}
\includegraphics{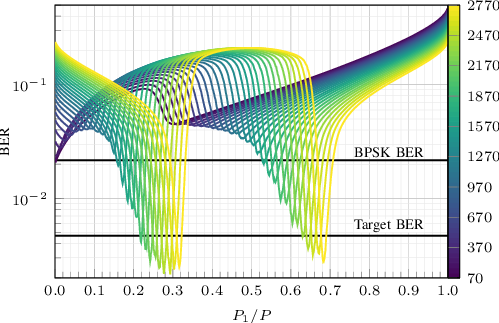}
\caption{BER for different rate and power split. Each curve corresponds to different $k_1 \in \left\{70, 170, \ldots, 2770\right\}$ information bits, $k_1 + k_2 = k$. Color corresponds to different values of $k_1$ represented by the colormap on the right of the figure. The horizontal axis corresponds to a linear fraction of transmit power allocated to the first code. SNR is $3.1$ dB.\label{fig:rate_split}}
\end{figure}

We have simulated the BER performance for $k_1 = 70, 170, \ldots, 2770 = k/2$ information bits allocated for the first code, considering a range of power fractions $P_1/P \in [0, 1]$ allocated to this code. The results are presented in~\Fig{fig:rate_split} for an SNR of $3.1$ dB. It is worth noting that for $k_1 = k/2$, the resulting plot is symmetrical with respect to the equal power split ($0.5$), indicating a minimum BER at $P_1^* \approx 0.32P$ and the same minima at $1-P_1^*$. Additionally, we observe that the BER is more sensitive to the power split than to the rate split. Decreasing the number of information bits in the first code leads to different optimal power allocation, resulting in asymmetric curves. The optimal point (using two regular LDPC codes) is as follows: the first code has $k_1 = 2470$ bits with power $P_1^* = 0.277$, while the higher-rate code ($k_2 = 3070$ information bits) has a larger power fraction ($P_2^* = 0.723$). This configuration yields error-free performance (FER $\ll 1$) for the second code.

The BER and FER of the optimized scheme are presented in~\Fig{fig:sim_results} as orange curves (marked as $2\times$ LDPC). The target BER is achieved at $3.05$ dB. It is worth noting that the resulting optimal power split is very close to the PAM-4 modulation. Specifically, the sum of two BPSK-modulated sequences with $P_1 = 0.2$ and $P_2 = 0.8$ will result in a set of transmitted points that match the PAM-4 constellation. Based on this idea, we constructed a single regular LDPC code with $k=5540$, $n_2=12000$, and simulated its performance using PAM-4 modulation. The number of channel uses in this configuration remains the same at $n=6000$. The resulting performance is very close to the superpositioning solution (see~\Fig{fig:sim_results}, green curves marked as PAM-4 LDPC). The target BER is achieved at $3.09$ dB. This raises the question of whether these schemes can be further improved and which of them will demonstrate better results.

\begin{table}
\caption{Optimized LDPC construction base graph, $k_b = 10$, factor $f=250$, distribution of check node degrees.\label{tab:proto_opt}}
\begin{tabular}{l|cccccc}
CN degree: &  1 &  2 &  3 &  8 & 13 & 14 \\
\hline
Count:     &  5 &  1 & 15 &  1 &  1 &  1 \\
\end{tabular}
\end{table}

% To answer this question, we performed the optimization of the LDPC code structure for each superpositioned component~\cite{Ebada2019zalup}. To improve the performance of the first code in the superpositioned scheme, we have optimized the parity check matrices of both codes and found that the second code (which actually defines the final bit error rate) mainly affects the final performance. To optimize the first code, we replaced the original regular parity check matrix with a lifted one. Then we optimized a base parity check matrix lifted with a factor $f=50$. We have fixed $k_1 = 2500$ in our optimization procedure (because the rate split has less effect on the performance than a power split) and tried to reduce the bit error rate as much as possible. We have shifted the target SNR point in accordance with the optimal power split we obtained in the regular LDPC case. By modifying the parity check matrices, we found that the minimum SNR at which the target BER is achieved becomes a $2.77$ dB (blue lines in~\Fig{fig:sim_results}. Our final base parity check matrix (having the size $14\times 24$) has a check node degree distribution very far from the regular one (see Table~\ref{tab:proto_opt}).

To address this question, we conducted an optimization of the LDPC code structure for each component in the superpositioned scheme~\cite{Ebada2019zalup}. To improve the performance of the first code, we optimized the parity check matrices of both codes and found that the second code, which determines the final bit error rate, has a greater impact on the overall performance. For the optimization of the first code, we replaced the original regular parity check matrix with a lifted one. Subsequently, we optimized a base parity check matrix lifted with a factor of $250$. In our optimization procedure, we fixed $k_1 = 2500$ (as the rate split has less effect on performance compared to the power split) and aimed to minimize the bit error rate. We adjusted the target SNR point according to the optimal power split obtained from the regular LDPC case. Through modifications of the parity check matrices, we found that the minimum SNR at which the target BER is achieved is reduced to $2.77$ dB (blue lines in~\Fig{fig:sim_results}). Our final base parity check matrix, with dimensions $14\times 24$, exhibits a check node degree distribution that significantly deviates from the regular one (see Table~\ref{tab:proto_opt}).

Then, we applied the same optimization technique to the PAM-4 scenario with a block length of $n_2 = 12000$, but we were unable to achieve a decrease in SNR. This can be attributed to the fact that uncoded PAM-4 modulation exhibits a very high input BER at the SNR of interest.

\section{Conclusion and future work}

\begin{table}
\caption{SNR at which the target BER $D$ has been achieved for $k=5540$ information bits, $n=6000$ channel uses\label{tab:final_numbers}}
\begin{tabular}{r|c|c}
SNR, dB & Coding scheme & Modulation \\   
\hline
$3.39$ & ER-optimized LDPC                           &           BPSK \\
$3.09$ & Regular LDPC, blocklength $2n$              &          PAM-4 \\
$3.05$ & $2\times$ superpositioned regular LDPC      & $2\times$ BPSK \\
$2.77$ & $2\times$ superpositioned ER-optimized LDPC & $2\times$ BPSK \\
$2.47$ & Sparse regression code, $M=2^8$             &       Gaussian \\
$2.16$ & Sparse regression code, $M=2^{10}$          &       Gaussian \\
\end{tabular}
\end{table}

In this paper, we have investigated the use of SPARCs in an error-reducing setup. SPARCs allowed us to achieve the same BER at a lower signal-to-noise ratio (SNR) by $1.23$ dB (see Table~\ref{tab:final_numbers}). Inspired by the performance of SPARCs, we also explored LDPC codes with superpositioning, as Gaussian modulation required by SPARCs may not be supported by hardware. The resulting gain in performance was reduced to $0.62$ dB. For future work, we plan to extend our study by considering a larger number of code components ($L > 2$) in the superpositioning scheme. Additionally, we will explore the use of polar codes with joint successive cancellation list decoding as proposed in~\cite{Marshakov2019Polar}. 
To better align with real-world wireless systems, we aim to investigate more realistic fading and multiple-input multiple-output (MIMO) channels.

\bibliographystyle{IEEEtran}
\bibliography{main}

\begin{thebibliography}{10}
\providecommand{\url}[1]{#1}
\csname url@samestyle\endcsname
\providecommand{\newblock}{\relax}
\providecommand{\bibinfo}[2]{#2}
\providecommand{\BIBentrySTDinterwordspacing}{\spaceskip=0pt\relax}
\providecommand{\BIBentryALTinterwordstretchfactor}{4}
\providecommand{\BIBentryALTinterwordspacing}{\spaceskip=\fontdimen2\font plus
\BIBentryALTinterwordstretchfactor\fontdimen3\font minus
  \fontdimen4\font\relax}
\providecommand{\BIBforeignlanguage}[2]{{%
\expandafter\ifx\csname l@#1\endcsname\relax
\typeout{** WARNING: IEEEtran.bst: No hyphenation pattern has been}%
\typeout{** loaded for the language `#1'. Using the pattern for}%
\typeout{** the default language instead.}%
\else
\language=\csname l@#1\endcsname
\fi
#2}}
\providecommand{\BIBdecl}{\relax}
\BIBdecl

\bibitem{Spielman1996erc}
D.~Spielman, ``{Linear-time encodable and decodable error-correcting codes},''
  \emph{IEEE Transactions on Information Theory}, vol.~42, no.~6, pp.
  1723--1731, 1996.

\bibitem{Polyanskiy2019jscc}
Y.~Kochman, O.~Ordentlich, and Y.~Polyanskiy, ``{A Lower Bound on the Expected
  Distortion of Joint Source-Channel Coding},'' in \emph{2019 IEEE
  International Symposium on Information Theory (ISIT)}, 2019, pp. 1332--1336.

\bibitem{Polyanskiy2018jscc}
------, ``{Ozarow-Type Outer Bounds for Memoryless Sources and Channels},'' in
  \emph{2018 IEEE International Symposium on Information Theory (ISIT)}, 2018,
  pp. 1774--1778.

\bibitem{Zamir2006jscc}
Z.~Reznic, M.~Feder, and R.~Zamir, ``{Distortion Bounds for Broadcasting With
  Bandwidth Expansion},'' \emph{IEEE Transactions on Information Theory},
  vol.~52, no.~8, pp. 3778--3788, 2006.

\bibitem{Verdu2013jscc}
V.~Kostina and S.~Verdú, ``{Lossy Joint Source-Channel Coding in the Finite
  Blocklength Regime},'' \emph{IEEE Transactions on Information Theory},
  vol.~59, no.~5, pp. 2545--2575, 2013.

\bibitem{Udo2011hamming}
B.~Müller, M.~Holters, and U.~Zölzer, ``{Low complexity Soft-Input
  Soft-Output Hamming Decoder},'' in \emph{2011 50th FITCE Congress - "ICT:
  Bridging an Ever Shifting Digital Divide"}, 2011, pp. 1--5.

\bibitem{Zhongfeng2018hamming}
Y.~Wang, J.~Lin, and Z.~Wang, ``{A Low-Complexity Decoder for Turbo Product
  Codes Based on Extended Hamming Codes},'' in \emph{2018 IEEE 18th
  International Conference on Communication Technology (ICCT)}, 2018, pp.
  99--103.

\bibitem{Polyanskiy2020graceful}
H.~Roozbehani and Y.~Polyanskiy, ``{Graceful degradation over the BEC via
  non-linear codes},'' in \emph{2020 IEEE International Symposium on
  Information Theory (ISIT)}, 2020, pp. 280--285.

\bibitem{Polyanskiy2019ldmg}
\BIBentryALTinterwordspacing
------, ``{Low density majority codes and the problem of graceful
  degradation},'' 2019. [Online]. Available:
  \url{http://arxiv.org/abs/1911.12263}
\BIBentrySTDinterwordspacing

\bibitem{Balitskiy2021}
G.~Balitskiy, A.~Frolov, and P.~Rybin, ``{Linear Programming Decoding of
  Non-Linear Sparse-Graph Codes},'' in \emph{2021 XVII International Symposium
  "Problems of Redundancy in Information and Control Systems" (REDUNDANCY)},
  2021, pp. 149--154.

\bibitem{Kschischang2017LDGMStairCase}
L.~M. Zhang and F.~R. Kschischang, ``{Low-Complexity Soft-Decision Concatenated
  LDGM-Staircase FEC for High-Bit-Rate Fiber-Optic Communication},''
  \emph{Journal of Lightwave Technology}, vol.~35, no.~18, pp. 3991--3999,
  2017.

\bibitem{Kschischang2018LDPCStairCase}
M.~Barakatain and F.~R. Kschischang, ``{Low-Complexity Concatenated
  LDPC-Staircase Codes},'' \emph{Journal of Lightwave Technology}, vol.~36,
  no.~12, pp. 2443--2449, 2018.

\bibitem{Li2020SparcsMAC}
T.~Li, Y.~Wu, M.~Zheng, D.~Wang, and W.~Zhang, ``{SPARC-LDPC Coding for MIMO
  Massive Unsourced Random Access},'' in \emph{2020 IEEE Globecom Workshops (GC
  Wkshps}, 2020, pp. 1--6.

\bibitem{Chamberland2023sparcs}
J.~R. Ebert, J.-F. Chamberland, and K.~R. Narayanan, ``{On Sparse Regression
  LDPC Codes},'' 2023.

\bibitem{Kschischang2014Staircase}
L.~M. Zhang and F.~R. Kschischang, ``{Staircase Codes With 6\% to 33\%
  Overhead},'' \emph{Journal of Lightwave Technology}, vol.~32, no.~10, pp.
  1999--2002, 2014.

\bibitem{Donoho2006CS}
D.~Donoho, ``{Compressed sensing},'' \emph{IEEE Transactions on Information
  Theory}, vol.~52, no.~4, pp. 1289--1306, 2006.

\bibitem{Donoho2010AMP}
D.~L. Donoho, A.~Maleki, and A.~Montanari, ``{Message passing algorithms for
  compressed sensing: I. motivation and construction},'' in \emph{2010 IEEE
  Information Theory Workshop on Information Theory (ITW 2010, Cairo)}, 2010,
  pp. 1--5.

\bibitem{Montanari2012Lasso}
M.~Bayati and A.~Montanari, ``{The LASSO Risk for Gaussian Matrices},''
  \emph{IEEE Transactions on Information Theory}, vol.~58, no.~4, pp.
  1997--2017, 2012.

\bibitem{BarronSparcs2019}
R.~Venkataramanan, S.~Tatikonda, and A.~Barron, ``{Sparse Regression Codes},''
  \emph{Foundations and Trends{\textregistered} in Communications and
  Information Theory}, vol.~15, no. 1-2, pp. 1--195, 2019.

\bibitem{Wang2019}
X.~Wang, S.~Cammerer, and S.~Ten~Brink, ``{Near-Capacity Detection and
  Decoding: Code Design for Dynamic User Loads in Gaussian Multiple Access
  Channels},'' \emph{IEEE Transactions on Communications}, vol.~67, no.~11, pp.
  7417--7430, 2019.

\bibitem{Ping2004}
L.~Ping, L.~Liu, K.~Wu, and W.~Leung, ``{Approaching the capacity of multiple
  access channels using interleaved low-rate codes},'' \emph{IEEE
  Communications Letters}, vol.~8, no.~1, pp. 4--6, 2004.

\bibitem{Ebada2019zalup}
\BIBentryALTinterwordspacing
A.~Elkelesh, M.~Ebada, S.~Cammerer, and S.~ten Brink, ``{Decoder-in-the-Loop:
  Genetic Optimization-based {LDPC} Code Design},'' 2019. [Online]. Available:
  \url{http://arxiv.org/abs/1903.03128}
\BIBentrySTDinterwordspacing

\bibitem{Marshakov2019Polar}
E.~Marshakov, G.~Balitskiy, K.~Andreev, and A.~Frolov, ``{A Polar Code Based
  Unsourced Random Access for the Gaussian MAC},'' in \emph{2019 IEEE 90th
  Vehicular Technology Conference (VTC2019-Fall)}, 2019, pp. 1--5.

\end{thebibliography}

\end{document}